\def\be{\begin{equation}}
\def\te{\end{equation}}
\def\bea{\begin{eqnarray}}
\def\nn{\nonumber}
\def\tea{\end{eqnarray}}

\def\ha{{1\over 2}}

\def\xib{\overline{\xi}}

\def\a{\alpha}
\def\b{\beta}
\def\d{\delta}

\def\g{\gamma}

\def\k{\kappa}

\def\m{\mu}

\def\n{\nu}

\def\L{\Lambda}
\def\O{\Omega}

\def\bb{\bibitem}
\def\mb{\overline{m}}

\def\oft{\overline{f}_2}

\documentstyle[aps]{revtex}
\input{epsf}
\begin{document}
\title{New quantum aspects of a vacuum-dominated universe}
\author{Leonard Parker\thanks{Electronic address: leonard@uwm.edu} and
Alpan Raval\thanks{Electronic address: raval@uwm.edu}\\
{\small \it Physics Department, University of
Wisconsin-Milwaukee, Milwaukee, WI 53201.}\\
WISC-MILW-00-TH-3}
\maketitle
\begin{abstract}
In a new model that we proposed, nonperturbative vacuum
contributions to the effective action
of a free quantized massive scalar field lead to a
cosmological solution in which 
the scalar
curvature becomes constant after a time $t_j$ (when the redshift
$z \sim 1$) that depends
on the mass of the scalar field and its curvature coupling. This
spatially-flat solution implies an accelerating universe at the present
time and
gives a good one-parameter fit
to high-redshift Type Ia supernovae (SNe-Ia) data, and the present age
and energy density of the universe. 
Here we show that the imaginary part of the nonperturbative curvature
term
that causes the cosmological acceleration, 
implies a particle production rate that agrees
with predictions of other methods and extends them to non-zero mass
fields.
The particle production rate is very small after the transition and is not
expected to alter the nature of the cosmological solution.  
We also show that the equation of state of our model 
undergoes a transition at $t_j$ from an equation of state dominated by
non-relativistic pressureless matter (without a cosmological constant) to
an effective 
equation of state of
mixed radiation and cosmological constant, and we derive the equation of
state of the vacuum. Finally, we explain why nonperturbative vacuum
effects of this ultralow-mass particle do not significantly change
standard early universe cosmology.\\

\noindent PACS numbers: 98.80.Cq, 04.62.+v, 98.80.Es
\end{abstract}
\section{Introduction}

Ever since the path-breaking prediction by Casimir \cite{casimir} that a quantized
field in its vacuum state will exert a force on nearby conducting plates,
and its confirmation in the laboratory \cite{casimirexpt}, the fundamental importance 
of the vacuum has been clear. The electromagnetic vacuum also manifests
itself through
other observable effects such as the Lamb shift \cite{lamb}, the anomalous
magnetic moment
of the electron \cite{schw}, and the anomaly-driven two-photon decay of
the pion \cite{adler}. The predicted
magnitudes of
these effects are obtained through renormalization of fields and coupling
constants, which absorb the infinities in a covariant manner. 
Renormalization methods have long been extended to quantized fields propagating
in the curved spacetime of general relativity \cite{utiyama} (see also
\cite{bandd} and references therein),
but predicted quantum vacuum effects of curved spacetime have been 
too small to be directly observed.
Quantum vacuum effects in curved spacetimes have also been explored in 
connection
with particle creation by the expansion of the universe and by black
holes \cite{parker,hawking}. In addition, vacuum effects of
self-interacting 
scalar fields have been invoked to produce inflation of the very early
universe \cite{guthlinde}.

As is well-known, observations of an accelerating expansion of the recent 
universe \cite{perl,riess}, together with other observations, such as those of 
small-angular-scale fluctuations of the cosmic microwave background radiation (CMBR), imply the
existence 
of a dark form of energy \cite{turner,sahni,frieman}. The present 
authors have argued that the recent acceleration of the 
universe is the first directly observable manifestation of quantum vacuum 
effects produced by the curved spacetime of general relativistic 
cosmology, and have presented a cosmological model, based on general relativity
and quantum field theory, that fits the current data \cite{parrav1,parrav2,parrav3}.
These quantum vacuum effects involve an ultralow-mass particle and are
nonperturbative in the curvature. They do not become significant until a
transition that occurs at about half the present age of the universe
(i.e., at a redshift $z \sim 1$.)

In the present paper, we explore new features of our model,
and their possible cosmological significance. In particular, we calculate
the creation rate of these ultralight particles resulting from the
effect of the spacetime curvature on the vacuum. 
The analytic result for the particle production rate during
the vacuum-dominated
stage predicted by our model reduces
to that obtained in the massless limit by other methods, and extends the 
result to the
massive case. This agreement adds confidence to the
validity of our effective action and its other predictions, such as that of a recent
acceleration of the expansion of the universe. {\it The predicted acceleration and the
particle
production arise from the real and imaginary parts, respectively, of the same term in the
effective action.}

We then find the effective temperature of
the particles created during the vacuum-dominated stage. 
This effective temperature at the present time would be very
small (only about $10^{-112}$ K), and is not expected to alter the
vacuum-dominated behavior. It should be kept in mind that the remnant of these particles
left over from the very early universe would have a much higher temperature today,
probably
of the order of 1 K.

The mass of the ultralight
scalar particle in our model is typical of masses expected for pseudo
Nambu-Goldstone
bosons (PNGB). In Ref. \cite{frieman} the PNGB mass scale arises
from the ratio of $m_{\nu}^2$ to $f$ ($m_{\nu} \sim 10^{-3}$ eV is a
neutrino mass and $f \sim 10^{18}$ GeV
is a global symmetry-breaking scale),
quantities that are {\it a priori}
independent of the 
present expansion rate of the universe. If the mass comes from a PNGB
mechanism, then this may introduce additional interaction terms into our
free-field model. Also, it is known that in a FRW universe, the graviton
field equation can be expressed in the form of two scalar field equations.
If the graviton has an ultralow mass, then we expect that it would lead to
the same cosmological consequences as the scalar particle discussed here.
Furthermore, such a low graviton mass appears to be consistent with
gravitational experiments and observations.

 The expansion
rate being related to the ultralow mass is not a coincidence in our model,
because
after the transition, such a relation between the mass of the particle
and the expansion rate remains true for a time of the order of the present 
age of the universe. In addition, the vacuum energy density remains within
an order of magnitude of the matter density for a time of the order of
the present age of the universe.  

The organization of this paper is as follows. In Section II, we briefly
summarize our model and the cosmological solution that arises from it. In
Section III, we discuss the particle production rate found from the
imaginary part of the effective action. In Section IV, we discuss the
fine-tuning problem in our model and in a model with cosmological constant
plus non-relativistic matter (referred to here as the $\L$-model). In
Section V, we derive
the
equation of
state of the vacuum in our model. In Section VI, we calculate the ratio of
total (i.e., matter plus vacuum contributions) pressure and total energy density for our
model and compare it to
that for a $\L$-model. This ratio as function of redshift is useful for future comparison to
observations. In Section VII, we discuss nonperturbative quantum
vacuum effects of this ultralow-mass field in the early universe, and
show that the standard
cosmological model is not significantly altered for times less
than about half the present age of the universe. Finally, in Section VIII,
we state our conclusions.  

\section{Nonperturbative vacuum energy effects}
We start with a brief summary of our non-perturbative vacuum energy model
and our previous results \cite{parrav1,parrav2,parrav3}.
Consider a free, massive
quantized scalar
field 
of inverse Compton wavelength $m$, 
and curvature coupling $\xi$. 
The effective action for gravity coupled to such a field is obtained by
integrating out the vacuum fluctuations of 
the field \cite{parrav1,parrav2,parrav3,partoms,jack} and renormalized
as in Ref. \cite{parrav1}. 
This effective action is 
the simplest one that gives
the standard trace anomaly in the massless-conformally-coupled limit, and
contains the nonperturbative sum (in arbitrary dimensions) 
of {\it all} terms in the 
propagator having at
least one factor of the  scalar curvature, $R$. It is given by
\cite{parrav1}
\bea
\label{effact1}
W &=& \int d^4 x \sqrt{-g} \k_o R + \frac{\hbar}{64 \pi^2} \int d^4 x
\sqrt{-g} \left[-m^4 \ln \mid \frac{M^2}{m^2} \mid \right. \nn
\\
& & \left. + m^2 \xib R \left(1-2\ln \mid \frac{M^2}{m^2} \mid
\right) - 2 f_2 \ln \mid \frac{M^2}{m^2} \mid + {3\over 2}
\xib^2 R^2 \right] \nn \\
& & + \frac{i\hbar}{64\pi} \int d^4 x \sqrt{-g}(M^4 + 2\overline{f}_2)
\theta (-M^2),
\tea
where $\k_o \equiv (16\pi G)^{-1}$
($G$ is Newton's constant), $\xib \equiv \xi -1/6$, and
\bea
M^2 &\equiv & m^2 + \xib
R \\
\overline{f}_2 &\equiv &(1/6)(1/5 -\xi)\Box R + (1/180)(R_{\a \b \g
\d}R^{\a \b \g \d} - R_{\a \b}R^{\a \b})\\
f_2 &\equiv &\overline{f}_2
+ (1/2)\xib^2 R^2. 
\tea
The renormalized cosmological constant has been set 
to zero in deriving the above effective action. The last term in the above
equation constitutes the
imaginary part of the effective action, implying particle production. We
will discuss the magnitude of the imaginary part in the next section. For
now, we focus on the real part of the effective action.
 
The trace of the Einstein equations, obtained by variation of
the real part of the  effective action with respect to the metric tensor
takes the
following form in a
Friedmann-Robertson-Walker (FRW) spacetime (in units such that $c=1$):
\bea
\label{one}
R + \frac{T_{cl}}{2\k_o} &=& \frac{\hbar m^2}{32 \pi^2
\k_o}\left\{\vphantom{\frac{m^2}{M^2}} (m^2 +\xib R)\ln 
\mid 1+\xib R m^{-2}\mid \right.\nn \\
& &-\left. \frac{m^2\xib
R}{m^2 +\xib R}\left(1+{3\over 2}\xib \frac{R}{m^{2}} + \ha \xib^2 
\frac{R^2}{m^{4}} (\xib^2 - (1080)^{-1}) +v \right)\right\}, 
\tea
where $T_{cl}$ is the trace of the energy-momentum tensor of  classical, 
perfect fluid matter
and $v \equiv 
(R^2/4 - R_{\m \n}R^{\m \n})/(180 m^4)$ is a curvature invariant
that vanishes in de Sitter space. As a result of including the
nonperturbative sum of scalar curvature terms in the effective action, the
right hand side of Eq. (\ref{one}) becomes large as $R \rightarrow
m^2/(-\xib)$. This resonance can occur even at low curvatures if $m^2$ is
sufficiently small, and is not displayed by a perturbative treatment of
$R$; as
can be seen by expanding Eq. (\ref{one}) in powers of $R/m^2$ and
keeping
a finite number of terms.

As noted earlier, $m$ is the inverse Compton wavelength of the
field. It is related to the actual mass of the field by $m_{\rm actual} =
\hbar m$. Equation (\ref{one}) above is nonperturbative in $R$ because it
contains terms that involve an infinite sum of powers of $R$. However, for a
sufficiently low mass, it is possible to treat 
$m_{\rm actual}^2/m_{Pl}^2 \equiv \hbar m^2/(16 \pi \k_o)$ (where $m_{Pl}$
is
the Planck mass) as a small
parameter and expand perturbatively in this
parameter.

For a sufficiently
low mass,
in an expanding FRW universe 
the quantum contributions to the Einstein equations become significant
at a time $t_j$, when the density of classical matter, $\rho_m$, 
has decreased to
a value given by
\be
\label{rhcond}
\rho_m(t_j) = 2\k_0 \mb^2,  
\te
where
\be
\mb^2 \equiv m^2/(-\xib).
\te
The time $t_j$ occurs in the
matter-dominated stage of the evolution.
Furthermore for $t > t_j$, the scalar curvature, $R$, 
remains constant to excellent approximation near the
value $\mb^2$. For $t<t_j$, the
quantum contributions to the Einstein equations are negligible and the
scale factor is that of a
matter-dominated FRW universe.
Then, equation (\ref{rhcond}) implies that, in a spatially flat universe
with line element $ds^2 = -dt^2 + a(t)^2(dx^2+dy^2+dz^2)$, one has
\be
t_j = (2/\sqrt{3})\mb^{-1},~~~~H(t_j) = \mb/\sqrt{3},
\te
where $H(t_j)$ is the Hubble constant at $t_j$.

The condition of the constancy of the
scalar curvature after $t_j$ leads to a solution for the scale factor 
that can be
joined, with continuous first and second derivatives (i.e., in a $C^2$
manner), to
the matter-dominated solution for $t < t_j$. The scale factor is
given by
\bea
\label{scfactor}
a(t) &=& a(t_j) \sqrt{\sinh\left.\left(\frac{t\mb}{\sqrt{3}} - \a
\right) \right/\sinh\left(\frac{2}{3}-\alpha\right)},~~~~t>t_j,\nn\\
&=& a(t_j)\left(\sqrt{3}\,\mb t/2\right)^{2/3},~~~~t<t_j,
\tea
with
\be
\a = 2/3 - \tanh^{-1}(1/2) \simeq 0.117.
\te
We would like to point out here that the above solution also satisfies, up
to terms of order
$m_{\rm actual}/m_{Pl}$, the one remaining
independent Einstein equation in a FRW universe, which can be taken to
be $G_{00} = (2\k_o)^{-1}T_{00}$, where 
$G_{\m \n}$ is the Einstein tensor and $T_{\m \n}$ is the energy-momentum 
tensor, including classical and quantum contributions. This
equation takes the form, with zero cosmological constant,
\bea
\label{zzcomp}
k_o G_{00} &=& \ha \rho_m - \frac{\hbar}{64\pi^2}\left\{ \frac{\xib
R_{00}}{m^2 +\xib R}\left(m^4 + 2 m^2\xib R + \frac{R_{\a \b}R^{\a
\b}}{90} +R^2 \left(\xib^2 -{1\over 270}\right)\right)\right.\nn \\
& & - 3
\xib^2
R\,R_{00}
 +\left.\vphantom{\frac{\xib}{R}} m^2 \xib
G_{00}\right\} 
- \frac{\hbar}{64\pi^2}\ln \left|1 + \frac{\xib R}{m^2}\right|
\left\{- \frac{m^4
g_{00}}{2}+ 2 m^2 \xib G_{00} \right.\nn
\\
& &-\frac{g_{00}R^2}{2}\left(\xib^2 +
\frac{1}{90}\right) 
+ \left.\frac{1}{90}g_{00}R_{\a \b}R^{\a \b} + 2\xib^2
R\,R_{00} - \frac{2}{45}{R_0}^{\a}R_{0\a}\right\}.
\tea
To verify that equation (\ref{scfactor}) is indeed a solution of the
above equation for $t>t_j$, we note that, when $R$ is very close to
the value $\mb^2$, the dominant terms in the right hand side of 
equation (\ref{zzcomp}) are those that have a factor of $m^2 + \xib R$ in
the denominator. Keeping these terms and substituting for the various
curvature quantities derived from equation (\ref{scfactor}), one finds
that equation
(\ref{scfactor}) satisfies
equation (\ref{zzcomp}) up to terms of order $m_{\rm actual}/m_{Pl}$. 

The solution in equation (\ref{scfactor}) corresponds to a universe that 
is accelerating
(i.e., has $\ddot{a} > 0$) for $t >
\sqrt{3}\mb^{-1}(\a + \tanh^{-1}(2^{-1/2})) \simeq 1.50\, t_j$. 
This solution
gives a good fit \cite{perl,riess} to the SNe-Ia data, for the mass range
\be
\label{mrange}
6.40 \times 10^{-33} \,{\rm eV} < \left(\frac{\mb}{h}\right) < 7.25 
\times 10^{-33}\, {\rm eV},
\te
where
$h$ is the present value of the Hubble constant, measured as a
dimensionless fraction of the value $100$ km/(s Mpc). 

The ratio of  matter density
to critical density at the present time, $\O_0$, is a function of
the single parameter $\mb/h$ and turns out to have the range $0.58 >
\O_0 > 0.15$ for the range of values of Eq. (\ref{mrange}). For
the same range of values, the age of the universe $t_0$ lies in
the range $8.10\,h^{-1}$ Gyr $<t_0<$ $12.2\,h^{-1}$ Gyr.  
These ranges for
$\O_0$ and $t_0$ agree with current observations \cite{bahcall}.

\section{Particle Production and Effective Temperature}

In this section, we consider the rate of particle production in the cosmological solution
discussed above. We find that the effective action from which our cosmological solution is
derived leads to a particle production rate that is consistent with that obtained using other
methods, and generalizes the other methods to the case of production of massive particles.
We also discuss the effective temperature of these particles.
 
The constant-$R$ solution after the transition at time
$t_j$ was obtained from a consideration of the Einstein equations based on variation of the
real part of
the effective action. The imaginary part of the effective action is
related to the probability of the production of at least one pair of
particles \cite{parker}, following Schwinger \cite{schwinger}. When the
imaginary part is small, this probability is given by
\be
P = 2\frac{{\rm Im}W}{\hbar}.
\te
>From Eq. (1), we obtain
\be
\label{imw}
\frac{{\rm Im} W}{\hbar} = (64\pi)^{-1}\int d^4x\sqrt{-g}(M^4 + 2\oft
)\theta(-M^2).
\te
When $M^2<0$ and $m=0$, the above formula reduces to that derived in Refs. \cite{zeldovich}
and \cite{dobado}. The derivation of the particle production rate in \cite{dobado} is based on 
the non-local effective action derived in \cite{vilkovisky}, which contains, in the $m
\rightarrow 0$ limit, terms of the form
$\ln (\Box )$. Similar terms were discussed in \cite{partoms2}. It is noteworthy that the
nonperturbative scalar curvature sum used here gives the same imaginary part, in
the limiting case of
zero mass, as the nonlocal effective action. The reason for the similar
results in the two cases was suggested in \cite{partoms2}, based on
renormalization group arguments \cite{rg}. However, these renormalization group arguments
are
generally reliable only in the large-$R$ limit. The partially summed
form of the effective action that we use here is also valid at small values of $R$, as
in the present universe.

For $m \neq 0$, Eq. (\ref{imw}) agrees, to leading order, with the
particle production
rate derived in \cite{birrell}. Since the imaginary and real parts of the effective action
arise out of the same term, the agreement of the imaginary part with that found in
the literature using other methods gives a further justification of our approximation.

For our cosmological solution, $M^2<0$ for all time, therefore the step
function in the above equation is equal to 1. The term $\oft \equiv (1/6)(1/5 - \xi)\Box
R + (1/180)(R_{\a \b \g \d}R^{\a \b \g \d} - R_{\a \b}R^{\a \b})$ can be
expressed as
\be
\oft = {1\over 6}\left({1\over 5} -\xi\right)\Box R + {1\over 120}C_{\a \b
\g \d }C^{\a \b \g \d} - {1\over 360} G,
\te
where $C_{\a \b \g \d}$ is the Weyl tensor, and $G$ is the Gauss-Bonnet
invariant
\be
G = R_{\a \b \g \d}R^{\a \b \g \d} - 4R_{\a \b}R^{\a \b} + R^2.
\te
In conformally flat spacetimes, such as the one under consideration, the
Weyl tensor vanishes. Therefore $\oft$ is a total derivative
term\footnote{$\Box R$ is obviously a total derivative, and in Robertson-Walker
spacetimes, $\sqrt{-g} G = 24 \frac{d}{dt}(\dot{a}^3/3 + k \dot{a})$, 
which is a total time derivative ($k =0, \pm 1$, corresponding to flat, closed or open
spatial sections).}, and $\int
d^4x\sqrt{-g}\oft$ only has boundary contributions. Since these boundary
contributions would vanish if the metric were static at the boundaries,
one expects that the $\oft$ term in the imaginary part of the
effective action does not correspond to real particle production \cite{parcarghese}. The
remaining term, $M^4$, will be used to estimate the probability of
particle production. 

\begin{figure}[hbt]
\centering
\leavevmode
\epsfysize=3.in\epsffile{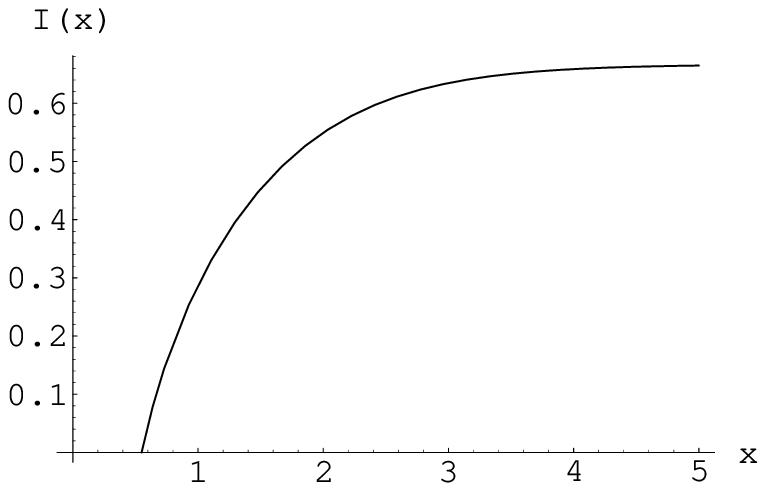}
\caption{A plot of ${\cal I}(x)$ versus $x= \overline{m} ct/3^{1/2} -
\alpha $. The 
graph begins at $t=t_j$ and asymptotes to a
value of $2/3$ as $x \rightarrow \infty$. The present time corresponds to
${\cal I}(x) \simeq 0.5.$}
\label{fig1nq}
\end{figure}

>From Refs. \cite{parrav1} and \cite{parrav2}, we find, for $t>t_j$,
\be
\label{m4}
M^4 \simeq (4320\pi)^{-2}
\left(\frac{m}{m_{\rm Pl}}\right)^4 \mb^4.
\te 
>From Eqs. (\ref{imw}) and (\ref{m4}), we obtain the probability per unit
proper volume, $p$, for production of at least one pair of particles
after time $t_j$, as
\be
\label{p}
p \simeq (32\pi)^{-1} (4320\pi)^{-2}\left(\frac{m}{m_{\rm Pl}}\right)^4
\mb^4 a(t)^{-3}\int_{t_j}^{t} dt' a(t')^3.
\te
To find the effective temperature of the particles produced, we compare
the above expression with the case when the particles are produced with a
thermal spectrum\footnote{Although the actual spectrum may not be thermal, the
high-frequency behavior of the spectrum is expected to be so 
\cite{parnature,parcincin,kandrup}.}.
For thermal production, we have \cite{dobado} 
\be
\label{pt}
p = \frac{\pi^2}{90}(k_B T)^3,
\te
where $T$ is the physical, or measured, temperature, and $k_B$ is
Boltzmann's constant. From Eqs. (\ref{p}) and (\ref{pt}), we obtain the
effective physical temperature of the particles produced, as
\be
\label{temp}
k_B T = \hbar \mb c^2
\left(\frac{90\sqrt{3}}{32(4320)^2\pi^5}\right)^{1/3}\left(\frac{m}{m_{\rm
Pl}}\right)^{4/3} {\cal I}(\mb c t)^{1/3},
\te
where we have now inserted appropriate factors of $c$, and ${\cal I}(\mb
ct)$ is the dimensionless function
\be
{\cal I}(\mb ct) = (\sinh x)^{-3/2}\int_{2/3 -\a}^{x}dy (\sinh y)^{3/2},
\te
with $x = \mb ct/\sqrt{3} - \a$. 
Note that the factor of $\hbar$ that appears in
Eq. (\ref{temp}) above is necessary because $\mb$ is an inverse length scale, rather than a
mass. 
A plot of ${\cal
I}(x)$ vs. $x$ is shown in Fig. \ref{fig1nq}. 
It is easy to verify analytically that ${\cal I}(\infty) = 2/3$. Also,
at the present time,
$t_0$ (defined as the time at which the Hubble constant has the value $65$
km/(s-Mpc)), one obtains ${\cal I}(\mb ct_0/\sqrt{3} - \a) \simeq 0.5$.

As shown earlier, a good fit to the SNe-Ia data is obtained when the mass
parameter $m_h$ has the value $6.93 \times 10^{-33}$ eV. With the rescaled Hubble
parameter, $h = 0.65$,
this gives a value $\mb = 4.5045 \times 10^{-33}$ eV. Also,
$m_{Pl} \simeq 1.2211 \times 10^{28}$ eV. One can reexpress the effective temperature
as
\be
T(\mb ct) \simeq 1.31 \times 10^{-112} {\rm K}\,\,{\cal I}(\mb ct)\,
\left(\frac{\mb}{4.5045\times10^{-33} {\rm eV}}\right)
\left(\frac{m}{4.5045\times10^{-33} {\rm eV}}\right)^{4/3}.
\te
If $m$ and $\mb$ are roughly equal to $4.5045 \times 10^{-33}$ eV, then the above temperature
corresponds to the temperature of the Hawking radiation emitted by a Schwarzschild black hole
of mass $\simeq 10^{138}$ g, which is about $10^{105}$ solar masses.

The fact that the one can associate a temperature with the created particles does not mean that
the vacuum possesses thermodynamic entropy. Indeed, from the equation of state of the vacuum, it
is straightforward to check that the change of entropy in a comoving volume of space $a^3 V$ is
given by $TdS = d(\rho_V a^3 V) + p_V d(a^3 V)$, which vanishes as a consequence of conservation
of vacuum energy.

\section{The fine-tuning problem}
In the absence of data that does not
distinguish between models for an accelerating universe, a criterion for 
the success of a given model is the lack of fine-tuning of fundamental
parameters. In a spatially
flat model with cosmological constant, $\Lambda$, and non-relativistic
matter ($\L$-model), the
fine-tuning problem is often understood as a coincidence between the 
matter density $\rho_m$
and the energy density associated with the cosmological constant, $\rho_{\Lambda}$, at
the present time. However,
it is
straightforward to show, in both our model and the $\L$-model, that the time
interval for which $\rho_m$ is, say, between $0.1$ and
$0.9$ times the critical density, $\rho_c$ 
is of the
order of the age of the universe. This is also the time interval for which  
$\rho_m$ and
the vacuum energy density\footnote{We will often refer to the vacuum
energy density as $\rho_V$, independent of the model under consideration. For the 
$\L$-model,
$\rho_V = \rho_{\L}$.} $\rho_{V}$ are roughly within an order of magnitude of each
other.
The calculation runs as follows. In our
model, $\O_0$ and $ht_0$ are the following functions of the single
parameter $m_h \equiv
\mb/h$ \cite{parrav2}:
\bea
\O_0 &=& (2.996 \times 10^6 {\rm Mpc})^2 m_h^2 \left(\frac{\sinh(cm_h
ht_0/\sqrt{3} - \a)}{\sinh(2/3 -\a)}\right)^{-3/2} \\
ht_0 &=& 3.26 \times 10^6 \frac{{\rm
yr}}{\rm Mpc}
\left(\frac{\sqrt{3}}{m_h}\right)\left(\tanh^{-1}(865.4\,{\rm Mpc}\,m_h) +
\a\right),
\tea
where $m_h$ is in units of Mpc$^{-1}$.
If $0.1 < \O_0 < 0.9$, then the above equations give $1.143 \times 10^{-3}
\,{\rm Mpc}^{-1} > m_h > 7.955 \times 10^{-4} \,{\rm Mpc}^{-1}$. For this
range of $m_h$ values, one then obtains $13.4\, {\rm Gyr} > ht_0 > 6.83
\,{\rm
Gyr}$. Thus the time interval for which $0.1 < \O_0 < 0.9$ is $(13.4 -
6.83)h^{-1}$ Gyr, or $6.57\,h^{-1}$ Gyr, which is of the order of the age
of the universe.

In the $\Lambda$-model, it is straightforward to show that $\O_0$ and $ht_0$
are functions of the single parameter $C \equiv \sqrt{6\pi G
\rho_{\L}}/(ch)$, which has dimensions of (time)$^{-1}$ (with the rescaled Hubble
parameter $h$ 
regarded as a dimensionless quantity). They are
\bea
\O_0 &=& 1 - {4\over 9}\left(C\,9.78 \, {\rm Gyr}\right)^2 \\
ht_0 &=& C^{-1} \tanh ^{-1}\left(\frac{2}{3}C\, 9.78\,{\rm Gyr}
\right) .
\tea
One similarly finds, for $0.1<\O_0<0.9$, that $12.49\, {\rm Gyr} > ht_0 >
6.75\, {\rm Gyr}$, giving a time interval of $5.74 h^{-1}$ Gyr, again of
the
order of the age of the universe.

The preceding arguments show that there is no ``coincidence''
problem in the sense of the present time being special in either model. 
However, in the case of the $\L$-model, an alternative statement of the
fine-tuning problem 
is that
there is no ``natural'' explanation, within elementary particle physics,
of a small non-zero value of $\rho_{\Lambda}$. The favored values of
$\rho_{\Lambda}$ are either $0$ or $m_{Pl}^4$, the latter value being at
discrepancy with observations by about $122$ orders of magnitude.

In our model, 
the question then is whether there may be a natural explanation
for the small value of the mass parameter $\mb \simeq H_0$. Frieman et
al. \cite{frieman} show that, for spin-0 pseudo Nambu-Goldstone bosons
(PNGB), a mass
scale $\mb$
of the order of the present value of
the Hubble constant can be generated from the neutrino mass $m_{\nu}
\simeq 10^{-3}$ eV and a global symmetry breaking scale (for example, the
scale for spontaneous breaking of the U(1) Peccie-Quinn symmetry) $f
\simeq 10^{18}$ GeV, via the combination $\mb \simeq m_{\nu}^2/f$. They
also show that this combination arises out of the coupling of such a
pseudo Nambu-Goldstone boson to neutrinos in a low-energy effective
theory. It therefore appears possible that the ultralight
spin-0 particle we consider in our model may be related to a PNGB,
although such a relation would involve additional self-interaction terms.

\section{Vacuum equation of state}
 Although the basic dynamical equations, (2) and (8),
incorporate non-trivial quantum effects, 
our model admits a remarkably simple description.  Indeed,
the scale
factor (\ref{scfactor}) may be used to find the total effective energy
density $\rho$ and pressure $p$ of vacuum plus matter by directly 
computing the Einstein tensor. 
We obtain, for $t>t_j$,
\bea
\label{rh}
\rho(t) &=& 2\k_0 G_{00} = (\k_0 \mb^2/2) \coth^2\left(t \mb/\sqrt{3}
- \a\right)\nn \\
& & = (3/2) \k_o \mb^2 \left(a(t)/a(t_j)\right)^{-4} +
(1/2)\k_o \mb^2\\
\label{p}
p(t) &=& 2\k_0 a(t)^{-2}G_{ii} = (\k_0 \mb^2/6) \coth^2\left(t \mb/\sqrt{3}
- \a\right) - (2/3)\k_0 \mb^2.
\tea
The effective equation of state for $t>t_j$ is therefore
\be
\label{eost}
p = (1/3)\rho - (2/3)\k_0 \mb^2,
\te
which is identical to the equation of state for a  classical model
consisting of radiation plus 
cosmological constant. 
In our
model, the equation of state of pressureless matter and the 
equation of state of quantum vacuum terms combine so as to appear as a sum
of radiation and cosmological
constant equations of state. Our model differs, even at the
classical level, from the usual mixed matter-cosmological constant
model because (i) for $t<t_j$ the effective cosmological constant
vanishes, 
and (ii) for $t>t_j$ vacuum contributions transmute the effective
equation of state into that of {\it radiation} (rather than
pressureless matter) plus cosmological constant; this surprising
metamorphosis is a result of the near-constancy of the scalar
curvature, which causes certain terms in $T_{\mu \nu}$ to take the
form of an effective cosmological constant term in Einstein's
equations. In a {\it general} spacetime, these terms do {\it not} have
the form of a cosmological constant term.

The equation of state for the quantum vacuum terms {\it alone} may be
inferred from equations (\ref{rh}) and (\ref{p}), and from the fact that
the density of pressureless matter is given by
\bea
\rho_m (t) &=& \rho_m(t_j)\,(a(t_j)/a(t))^3  \nn \\
&=& 2\k_0 \mb^2 \left(\sinh(2/3 -
\a)\left/\sinh(t\mb/\sqrt{3}-\a)\right)\right.^{3/2},
\tea
where equations (\ref{rhcond}) and (\ref{scfactor}) have been used 
to arrive at the second equality. The quantum vacuum energy density
$\rho_V$ and pressure $p_V$ then follow, for $t>t_j$, as
\bea
\rho_V(t) &=& \rho(t) - \rho_m(t) \nn \\
& &= \frac{\k_0
\mb^2}{2}\left[\coth^2\left(t \mb/\sqrt{3} - \a\right) - 4
\left(\frac{\sinh(2/3 -
\a)}{\sinh(t\mb/\sqrt{3}-\a)}\right)^{3/2}\right]\\
p_V(t) &=& p(t),     
\tea
with $p(t)$ given by equation (\ref{p}). The above equations
show that for $t>t_j$ the vacuum energy density is positive, while the
vacuum pressure is negative. As stated earlier, the vacuum
terms are negligible for $t<t_j$.

As a consequence of equations (13) and (14), we find that the equation of
state for the vacuum is
\be
\label{eosv}
\rho_V = 3 p_V + 2\mb^2 \k_0\left[1-\left(1+
2p_V/(\k_0\mb^2)\right)^{3/4}\right].
\te
This vacuum equation of state joins continously to the equation of state
$\rho_V = p_V = 0$ at $t=t_j$ and is asymptotic to the
pure cosmological constant equation of state $\rho_V =
-p_V$ as $t \rightarrow \infty$. Equation (\ref{eosv})  
is parametrized by the single parameter, $\mb$, and
is different from the equation of state of a pure cosmological
constant. However, Eq. (\ref{eosv}) holds only in the case of
Robertson-Walker symmetry. Any inhomogenous perturbation of the matter
content
or the metric in our model would change the effective equation of state of
the vacuum, since such perturbations would change the form of the
right-hand-sides of Eqs. (2) and (8). This feature is a complication when
one analyzes the evolution of small perturbations in the theory. However,
we expect that the terms of Eq. (\ref{eosv}) may still dominate the
equation of state for the perturbations.

\section{Ratio of pressure and energy density}

In this section, we derive the ratio of total pressure $p$ and total matter density $\rho$ in
our model,
as a function of redshift. This quantity, which we call $w(z) \equiv p/\rho$, is a
useful one for
comparison with future observations. We also compare $w(z)$ in our model with the
same quantity in a model with non-relativistic matter plus cosmological constant.

>From Eqs. (24) and (25) above, we find that, for $z<z_j$, 
\be
\label{w1}
w(z) = {1\over 3} - {2\over 3}\frac{\k_o \mb^2}{\rho},
\te
with
\be
\label{dens1}
\rho (z) = \k_o \mb^2 \left\{ {3\over 2} \left(\frac{1+z}{1+z_j}\right)^4 + {1\over 2}
\right\}.
\te
The redshift at which the transition occurs, $z_j$, is given by
\be
\label{zj1}
1+z_j \equiv \frac{a(t_0)}{a(t_j)} = \sqrt{\frac{\sinh\left(ct_0 \mb /\sqrt{3} -
\a\right)}{\sinh \left(2/3 - \a\right)}},
\te
where we have inserted an appropriate factor of $c$. Using the value of
$\a$ from Eq. (7), and
the present value of the Hubble constant,
\be
H_0 = \frac{c \mb}{\sqrt{12}} \coth \left(\frac{c t_0 \mb}{\sqrt{3}} - \a\right),
\te
one can reexpress Eq. (\ref{zj1}) as
\bea
\label{zj2}
1+z_j &=& 
\left(\frac{4 H_0^2}{c^2 \mb^2} - {1\over 3}\right)^{-1/4} \nn \\
&=& \left\{0.3788 \,\left(\frac{6.93 \times 10^{-33} {\rm eV}}{m_h}\right)^2 - {1\over
3}\right\}^{-1/4},
\tea
where we have substituted for the numerical value of the speed of light, $c$, in the second
equality.

Combining Eqs. (\ref{w1}), (\ref{dens1}) and (\ref{zj2}), we obtain
\be
w(z) = {1\over 3} - {4\over 3} \left\{1.1364 (1+z)^4 \left(\frac{6.93 \times
10^{-33} {\rm eV}}{m_h}\right)^2 - (1+z)^4 + 1 \right\}^{-1},
\te
for $z<z_j$. For $z>z_j$, our model is entirely 
matter-dominated, therefore $p = w(z) = 0$.

On the other hand, in a spatially flat model with non-relativistic matter plus cosmological
constant, it is
straightforward to show that
\be
w(z) = - \left\{ 1 + (1+z)^3 \left(\O_{\L}^{-1} - 1\right) \right\}^{-1},
\te
where $\O_{\L}$ is the ratio of the energy density in the cosmological constant and the
critical density.

\begin{figure}[hbt]
\centering
\leavevmode
\epsfysize=3.in\epsffile{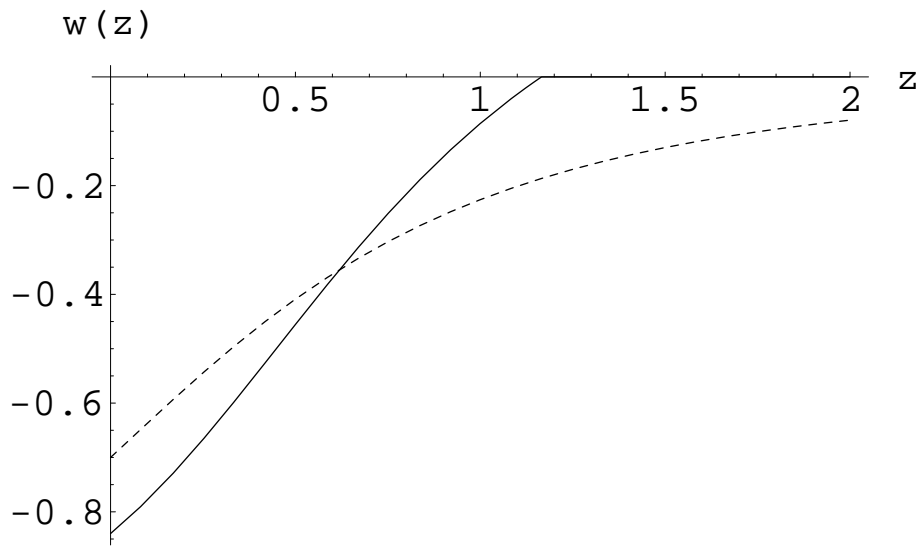}
\caption{A plot of $w(z)$ versus redshift $z$, in our model (solid curve) 
and a mixed matter
plus cosmological constant model (dashed curve). The relevant parameter 
values are $m_h =
6.93 \times 10^{-33}$ eV (solid curve), and $\O_{\L} = 0.7$ (dashed curve).}
\label{fnq1}
\end{figure}

A plot of $w (z)$ for both models discussed above is shown in Fig. 
\ref{fnq1}, with
representative values of $m_h = 6.93 \times 10^{-33}$ eV (in our model), and $\O_{\L} = 0.7$
(in the non-relativistic matter plus cosmological constant model). Future experimental data
should be able to distinguish between these models.

Noting that the equation of state of non-relativistic matter would give
$w(z) =0$, and the equation of state of a pure cosmological constant would
give $w(z) =-1$, we see from Fig. \ref{fnq1} that the redshift interval
for which the effects of non-relativistic matter and vacuum energy are
both significant is $\Delta z \simeq 2$. This range is a small one
compared to the full redshift range, which is infinite. This feature is
often taken to comprise an alternative statement of the
``coincidence'' problem outlined in the previous section. However, as we
showed there, the small redshift range corresponds to a large range in
time. The discrepancy between posing the ``coincidence'' problem in terms
of time range and redshift range reflects the subjective nature of this
problem. We reiterate that the most reasonable way to pose the fine-tuning
problem is in terms of a fundamental explanation for the smallness of $\L$
or $m_h$.

\section{Nonperturbative effects of an ultralow mass field in the early
universe}
In this section, we consider the possible quantum vacuum effects of an
ultralow mass field ($m \sim 10^{-33}$ eV) in the early, post-inflationary
universe. We first note that quantum vacuum effects become
significant whenever the ``resonance'' condition is satisfied, i.e.,
whenever the
trace of the
classical stress tensor $T_{\rm cl} \equiv -\rho_{\rm cl} + 3p_{\rm cl}$
decreases below the value $2\k_o \mb^2$ (see the arguments leading from
Eq.
(5) to Eq. (6) for details). Consider now the evolution of $T_{\rm cl}$
from early inflation to the present time. During early inflation, $T_{\rm
cl}$ has a large negative value, i.e., $\mid T_{\rm cl}\mid  \gg 2\k_o
\mb^2$. After the reheating process at the end of inflation, the universe
is expected to be radiation-dominated, with $\mid
T_{\rm cl} \mid $ decreasing to a small value. As the universe
expands, and enough matter becomes
non-relativistic, $\mid T_{\rm cl} \mid $  increases smoothly to a maximum
value much greater than $2\k_o \mb^2$.
As the universe
expands further, it enters a matter-dominated stage. 
During this stage, $\mid T_{\rm cl} \mid $ decreases again,
until about half the
age of the universe ($z \sim 1$), when it passes through
 the value $2\k_o \mb^2$, and quantum vacuum effects begin to dominate,
leading to the present acceleration. However, as described above, there
are two early stages of the evolution during which $\mid T_{\rm cl} \mid $
may pass through the value $2\k_o \mb^2$, namely, during the reheating
from inflationary expansion to radiation-dominated expansion, and when
$\mid T_{\rm cl} \mid$ is increasing during the
radiation-dominated stage of the expansion of the universe. 
What role do quantum vacuum
effects of this ultralow-mass particle play during these two early
stages?

To answer this question, we consider two possibilities for the value of
$\mid T_{\rm cl} \mid $ during the radiation-dominated stage: (a) there is
sufficient non-relativistic matter at the end of post-inflationary
reheating so that $\mid T_{\rm cl} \mid $ is always greater than $2\k_o
\mb^2$ during the radiation-dominated stage, or (b) there is not enough
non-relativistic matter, and the condition, $\mid T_{\rm cl} \mid < 2\k_o
\mb^2$, is satisfied at some time  during
the radiation-dominated stage.

If possibility (a) occurs, then the ``resonance'' condition is never
satisfied during the reheating from inflationary to radiation-dominated
expansion or the subsequent evolution from radiation-dominated
to matter-dominated expansion. Thus these quantum vacuum effects of 
an ultralow-mass particle remain
negligible in the early universe and in the matter-dominated stage,
until the universe transits to vacuum-dominance at about half its
present age, as
described in Section II.

If possibility (b) occurs, then a resonance of the type discussed here
will occur during the reheating from inflationary expansion to
radiation-dominated expansion, and will force $R$ to remain nearly equal
to $\mb^2$ during the early part of the 
radiation-dominated stage. During such a
stage, the scale factor would be $a(t) \propto
\sqrt{\sinh(t\mb/\sqrt{3})}$.
However, $t\mb \ll 1$ during this stage, and the scale
factor, to excellent approximation, exhibits radiation-dominated behavior,
i.e., $a(t) \propto t^{1/2}$. Once enough non-relativistic matter has
accumulated, $\mid T_{\rm cl} \mid $ increases above the value $2 \k_o
\mb^2$ while the universe is still radiation-dominated. At this point, 
quantum effects become negligible, as can be seen from Eq. (5). 
Once there is sufficient non-relativistic matter, 
the exit from vacuum-dominance must 
occur\footnote{This exit is analogous to the time-reversed 
evolution of the present vacuum-dominated
stage through the transition at time $t_j$.}. 
Eventually, the universe transits to the matter-dominated stage.
Much later, at time $t_j$,
when the condition $\mid T_{\rm cl} \mid = 2\k_o \mb^2$ is again
satisfied, the universe transits to
vacuum-dominance, giving rise to the present acceleration. However,
because the density of non-relativistic matter (or equivalently, $\mid
T_{\rm cl} \mid $) is now decreasing, the universe remains
vacuum-dominated
for all time.

The above arguments show that, even if these
ultralow-mass nonperturbative quantum vacuum effects become
significant at early times in the evolution of the universe, the
eventual transition to the present vacuum-dominated stage is not 
affected and the early universe dynamics is not significantly altered 
beyond the standard cosmological model.   
     
\section{Conclusions}
The cosmological model that we investigate here is based on nonperturbative
vacuum effects of a quantized noninteracting massive scalar field. This model
is the simplest one exhibiting these nonperturbative quantum effects. It is
based on the extension of quantum field theory to curved spacetime, a subject
that has been extensively developed and has given rise to fundamental
advances in black hole physics and cosmology.
 
In the present paper, we have studied the particle production resulting from
the imaginary part of the same term that, in earlier work, we showed would
cause, through its real part, an acceleration of the universe consistent with
that inferred from recent supernovae observations and other cosmological
data. We find that the predicted rate of particle production during the
recent vacuum-dominated era is the same as that found by other methods. It
has been shown \cite{dobado} that this production rate also agrees with that
found by the Bogolubov transformation technique \cite{parker,parcincin}. This
agreement of the particle production rate given by the imaginary part of our
effective action, adds confidence to the validity of the acceleration
resulting from the real part of the same term in the effective action.

The density and effective temperature of these ultralight particles created
after the transition to vacuum dominance is negligible compared to the matter
and radiation already present. However, the production of these particles in
the early universe would leave a remnant of such particles similar to the
remnant of gravitons created near the big bang. The mass of these particles
is such that they may contribute a component to dark matter that is
sufficiently cold to be gravitationally concentrated in the vicinity of
galaxies. The possibility that these may contribute to extended galactic
halos is worth exploring, as is the effect, if any, of these particles on
big-bang nucleosynthesis in the early universe.

The evolution of the scale factor, $a(t)$, in our cosmological solution depends on a
single parameter related to the mass $m$ of the particle. Cosmological data give a
value for $m$ of order $10^{-33}$ eV. This mass is of the same order of magnitude as
the mass of a pseudo-Nambu-Goldstone boson, obtained by independent particle physics
considerations \cite{frieman}. As discussed in Sec. IV, this fact may be
significant with regard to the question of fine-tuning.

We calculated the effective equation of state during the vacuum-dominated era
in our model, and gave the function $w(z) \equiv p/\rho$. The effective
equation of state of vacuum plus matter during the vacuum-dominated stage
turns out to be the same as that of a classical model with radiation plus
cosmological constant, although the actual matter content is mainly
non-relativistic.

We showed that these nonperturbative vacuum effects of an ultralow
mass field would not significantly alter standard cosmology at early
times. The
possible existence of much more massive scalar fields, for which these
nonperturbative vacuum effects would be significant in the very early
universe, may
enhance inflation. In addition, the associated particle production may help
reheat the universe during the exit from early inflation. It is possible that
processes, such as spontaneous symmetry breaking (caused, for example, by a
self-interaction), may cause the vacuum expectation value of the field to
become non-zero, while its fluctuations are reduced. In the present
non-interacting model, such processes are absent. The effect of 
interactions, such as those arising from a PNGB, on the observational
predictions of the present model (involving
an ultralight mass) is also worth exploring.

Also worth noting is that quantum vacuum terms would become dominant in
regions of low average density
(i.e., $\rho_m 
< 2\mb^2 \k_o$) earlier than in regions of high average density. This would alter the
evolution of  density inhomogeneities that existed during the early matter-domnated stage of
the expansion. This may
eventually provide another observational test of our model. 

Our present model may soon be observationally tested relative to other models
through observations of the small-scale CMBR fluctuations. These fluctuations
are determined by the function $w(z)$ (given in Sec.~VI). Non-zero
spatial curvature would, of course, affect the predicted shape of the
CMBR fluctuation spectrum, as would the addition of
interactions to our present free-field model.\\
\bigskip\\
\noindent {\bf Acknowledgements}\\

\noindent The authors would like to thank C. T. Hill and E. W. Kolb for
helpful
comments. This work was supported by NSF grant PHY-9507740.


\end{document}